\documentclass[letterpaper, 10 pt, conference]{ieeeconf}  

\IEEEoverridecommandlockouts                              

\usepackage{amsmath, amsthm, amsfonts, amssymb,color}
\usepackage[ruled,vlined,noend]{algorithm2e}
\usepackage{blindtext}
\usepackage{graphicx}
\usepackage[utf8]{inputenc}
\usepackage{mathtools}
\usepackage{tikz, cite}
\usepackage[font=footnotesize,labelformat=simple]{subcaption}
\usepackage{tabu}

\newtheorem{definition}{Definition}
\newtheorem{example}{Example}
\usepackage{epsfig} 
\usepackage{psfrag}
\usepackage{mathrsfs}

\usepackage{multirow}
\usepackage{array,booktabs,arydshln,xcolor}
\usepackage{soul}
\usepackage{mathtools}
\usepackage{nccmath}
\usepackage{caption}
\usepackage{nccmath}
\usepackage{comment}
\usepackage{pdfpages}

\usepackage{comment}
\renewcommand{\vec}[1]{\boldsymbol{#1}}
\usepackage{float}
\usepackage{graphicx}
\usepackage{epstopdf}
\usepackage{color}
\usepackage{verbatim}
\usepackage{tikz,times}
\usepackage{xcolor}
\usepackage{lipsum}
\usepackage{makecell}
\let\labelindent\relax

\usepackage{enumitem}
\usepackage{multirow}

\newtheorem{assumption}{\textbf{Assumption}}

\usepackage{stfloats}

\DeclareMathOperator*{\argmin}{arg\,min}  

\title{\LARGE \bf
Learning-Aided Warmstart of Model Predictive Control in Uncertain Fast-Changing Traffic
}

\author{Mohamed-Khalil Bouzidi$^{1,2}$, Yue Yao$^{1,2}$, Daniel Goehring$^{1}$, Joerg Reichardt$^{2}$
\thanks{$^{1}$ is with the Free Universitity of Berlin, Germany }
\thanks{{\tt\small \{firstname.lastname@fu-berlin.de\}}}
\thanks{$^{2}$ is with Continental AG}
\thanks{{\tt\small \{firstname.lastname@continental.com\}}}
\thanks{This work received support by the German Federal Ministry for}
\thanks{Economic Affairs and Climate Action within the project KI Wissen.}
}

\begin{document}

\maketitle
\thispagestyle{empty}
\pagestyle{empty}

\begin{abstract}
 Model Predictive Control lacks the ability to escape local minima in nonconvex problems. Furthermore, in fast-changing, uncertain environments, the conventional warmstart, using the optimal trajectory from the last timestep, often falls short of providing an adequately close initial guess for the current optimal trajectory. This can potentially result in convergence failures and safety issues. Therefore, this paper proposes a framework for learning-aided warmstarts of Model Predictive Control algorithms. Our method leverages a neural network based multimodal predictor to generate multiple trajectory proposals for the autonomous vehicle, which are further refined by a sampling-based technique. This combined approach enables us to identify multiple distinct local minima and provide an improved initial guess. We validate our approach with Monte Carlo simulations of traffic scenarios.
\end{abstract}
\section{Introduction} \label{sec:Intro}

Model Predictive Control (MPC) has established itself as a popular technique in Motion Planning and Control for autonomous driving. This is attributed to its inherent capability to simultaneously account for collision constraints, dynamic feasibility, actuator constraints, and comfort criteria, enabling the generation of optimal trajectories \cite{micheli, siebenrock, shi}.  A notable variant that we also use is Model Predictive Contouring Control (MPCC) \cite{liniger, brito1, lyons}. It generates consistent lateral and longitudinal control signals and does not require a separate desired velocity specification. However, due to constrained computational resources, MPC relies on local optimization, employing simple models and limited planning horizons, potentially resulting in suboptimal or locally optimal (short-term) solutions. Conversely, learning-based approaches can excel where MPC falls short e.g., in efficiency and adaptability in complex tasks, without needing physical models\cite{aradi, kendall}. However, they face challenges in interpretability and reliability, especially in unexplored corner cases. This can potentially lead to hazardous behavior, hindering their suitability for critical applications. Hence, due to their complementary attributes, several methods propose approaches to combine MPC with learning-based approaches. 

Learning-based MPC can be broadly categorized into two groups. The first group employs a learning-based system to substitute or enhance components of MPC. Simplest are approaches that learn the weights of the cost function\cite{gros, zarrouki}, as these significantly impact MPC performance and can be challenging to tune manually. A similar technique is cost shaping \cite{brito2, tamar} which adjusts the cost function at each time step, mitigating MPC's limitation in finding only short-term optimal solutions. Other methods learn the state-space model or parts of it \cite{Koller,Berberich, Hewing} to handle unknown or complex dynamics.
\begin{figure}[t]
\centering
    \includegraphics[width=0.48\textwidth]{"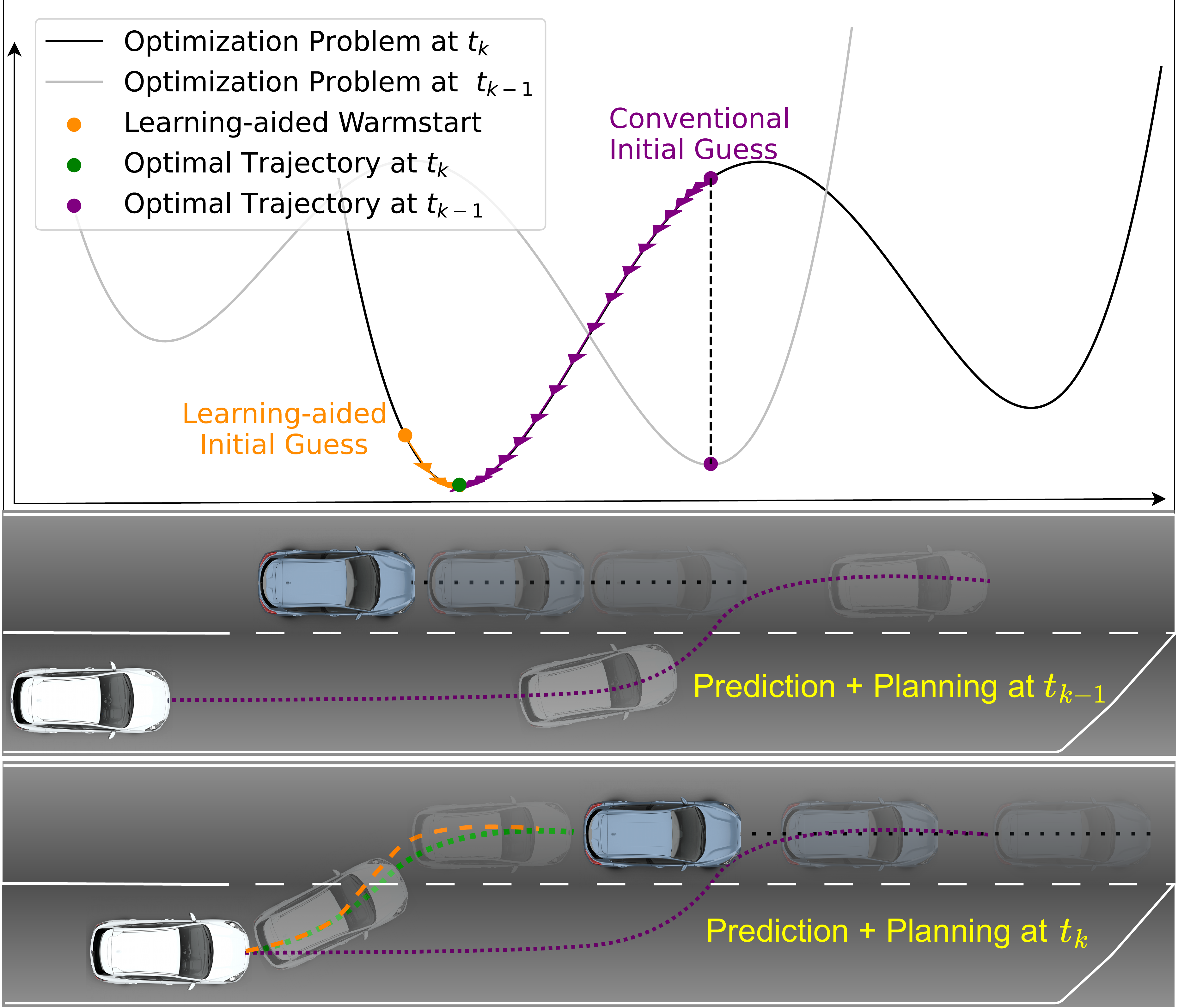"}
    \caption{Example where our warmstart improves convergence quality compared to warmstarting with the solution of the last timestep $t_{k-1}$ due to change of the optimization problem (changing  traffic participant behavior prediction)}
    \label{fig:P1}
\vspace{-1.5em}    
\end{figure}
The second group learns high-level policies where the trajectory is further refined with low-level MPC. Methods such as \cite{brito3, song} provide high-level plans as a reference to the MPC. Similarly, the predictive safety filter \cite{Wabersich, Tearle} evaluates constraint satisfaction of the trajectory of the learned system, potentially generating an output that minimizes the discrepancy from it while adhering to constraints.

Our approach of a learning-based warmstart also falls into this group, together with \cite{mansard, kabutan, natarajan, lembono}. Here, the learned system offers an initial guess to the MPC optimizer, which is then further optimized by the MPC. This concept is particularly compelling given the inherent limitations of Local Optimizers/MPC, that become apparent in the context of autonomous driving in complex scenarios. The first well-known deficiency of the local optimizer is that if the initial guess is far from the optimum, many steps are needed until it converges, or the optimization may not converge at all. The strategy of MPC to provide an initial guess is to use the optimal trajectory, which was calculated in the last timestep, assuming little change between the previous and current timestep. This strategy fails in uncertain and rapidly changing environments where the optimization problem can vary a lot between each timestep (s. Fig. \ref{fig:P1}). For instance, due to unknown intentions of human drivers, predictions of how traffic participants act may vary significantly between timesteps. These abrupt changes can lead the optimizer to struggle to recover or find a proper solution in time, potentially resulting in fatal behavior where e.g., collisions cannot be avoided. In the event of optimizer failure, a common approach is to use the same control input as in the last timestep. However, when the environment is changing rapidly, the scene can change even more in the time step after and now using the solution from two timesteps ago only exacerbates the problem.

The second deficiency of only being able to find a local optimum is especially problematic in dense traffic with (moving) obstacles. These obstacles are generally the cause for non-convex problems with multiple local minima. Some of these minima lead to undesired behavior, such as overly conservative driving or peculiar overtaking maneuvers. This problem is often mitigated by decomposing the planning problem into a global and a local planner \cite{svenstrup, li}.  In this setup, the global planner generates a rough trajectory with significant simplifications for real-time feasibility,  potentially sacrificing optimality. Also, topology-based planners such as \cite{degroot, rosmann, yi} can address this weakness, from which we adopt the concept of homotopy classes. But, these planner also do not address the previously mentioned weakness of conventional warmstarting in fast-changing environments that our method tackles. 

However, previous works for learning-based  warmstarting \cite{mansard, kabutan, natarajan, lembono} are mainly designed for simple repeating tasks. For example, they do not consider constraints, especially moving obstacles, or are trained for a limited number of self-generated scenarios (which additionally require retraining when  the weights of the MPC cost function change).  

The main contributions of our work are summarized as:
\begin{itemize}[leftmargin=*]
    \item Designing a Motion Planner based on  Model Predictive Contouring Control with Artificial Potential Fields
    \item Developing a learning-aided warmstart strategy which improves convergence quality in fast-changing unknown scenarios and helps to prevent undesired local minima leveraging the concept of homotopy classes
    \item Devising a time-efficient framework with a novel trajectory refinement process which makes arbitrary multimodal trajectory predictors learned on real-world datasets easily deployable.
\end{itemize}

\section{Baseline Model Predictive Contouring Control} \label{sec:MPCC}
\begin{figure}[t!]
\vspace{0.6em} 
\centering
    \includegraphics[trim = 8mm 0mm 12mm 12mm, clip, width=0.475\textwidth]{"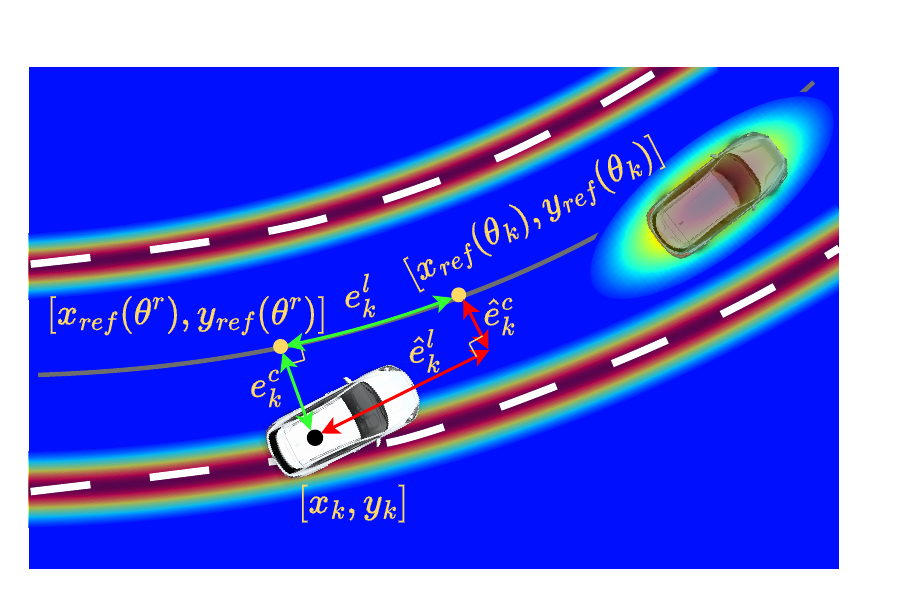"}
    \caption{Illustration of the lag error $e^l_k$, contouring error $e^c_k$ (where $\theta^r$ is the real arclength) and potential field on obstacles and lane markers}
    \label{fig:P2}
    \vspace{-0.8em} 
\end{figure}

Consider an arbitrary traffic scene with $O$ traffic participants $o \in \{0,..,O-1\}$ in which an autonomous vehicle (AV), denoted as $o= 0$, needs to plan and execute a safe trajectory. A reference path and the map $M_r$ with road boundaries are given by sets of waypoints $p_l = \{(x^l_j,y^l _j)\}^{K}_{j=0}$ with $ l \in \{ref, lb, rb\}$ possibly provided by a high-level route planner. The reference path is parameterized by the arclength $\theta$ and augmented by the path orientation $\psi_{ref}$ and the distance to the left and right road boundary $d_{lb}$, $d_{rb}$, $\mathcal{P}_{ref}: [0,\theta_{max}] \to \mathbb{R}^2 \times [0, 2\pi], \theta \mapsto (x_{ref}(\theta), y_{ref}(\theta), \psi_{ref}(\theta), d_{lb}(\theta), d_{rb}(\theta))$ where $\theta_{max}$ is the maximum arclength. 

We model the motion of the AV by a differential equation $\dot{\vec{z}}(t) = f(\vec{z}(t), \vec{u}(t))$ using the kinematic bicycle model:
\begin{equation}
    \dot{\vec{z}} = \left[ v \cos(\psi), \ v \sin(\psi), \ v \frac{\tan(\psi)}{l}, \ a, \ j, \ \dot{\delta} \right]^\top
\end{equation}

where $\vec{z} = [x,y,\psi, v, a, \delta]^\top$ is the state vector, $\vec{u} = [j, \dot{\delta}]^\top $ is the control input vector and the velocity, acceleration, steering angle, jerk, steering angle rate, and wheelbase are denoted as $v, a, \delta, j, \dot{\delta}, l $, respectively.

For the MPC-formulation, the dynamic model is discretized to $\vec{z}_{k+1} = f(\vec{z}_k, \vec{u}_k)$ with the sampling time $T_s$.
The MPCC aims to maximize path progress while minimizing path error, balancing between the two objectives. For that, we approximate the arclength (i.e. progress on the path)  $\theta_{k}$, the lag error $\hat{e}^l_k$ and the contouring error $\hat{e}^c_k$ (s. Fig \ref{fig:P2}):
\begin{equation}
\begin{aligned}\label{eq:2}
& \theta_{k+1} = \theta_{k} + v^p_k T_s \\  
\end{aligned}
\end{equation}

\begin{equation}
\begin{aligned}
& \left[\begin{array}{cr}
\hat{e}^c_k \\
\hat{e}^l_k 
\end{array}\right]
=\left[\begin{array}{cr}
\sin (\psi_{ref} \left(\theta_k\right)) & -\cos (\psi_{ref} \left(\theta_k\right)) \\
-\cos (\psi_{ref} \left(\theta_k\right)) & -\sin  (\psi_{ref} \left(\theta_k\right))
\end{array}\right]
\Delta \vec{p}_{ref}\nonumber
\end{aligned}
\end{equation}

where $\Delta \vec{p}_{ref} = [x-x_{ref}\left(\theta_k\right), \ y-y_{ref}\left(\theta_k\right)]^\top $ and $v^p_k$ is the virtual speed on the path. Eq. \ref{eq:2} is augmented to the dynamic model, i.e. $v^p_k$ is an additional control input and  $\theta_k$ a further state.
This is utilized to define the running cost:
\begin{equation}\label{eq:3}
    J_{k} = \left[\begin{array}{cr}
\hat{e}^c_k \\
\hat{e}^l_k 
\end{array}\right]^\top Q 
\left[\begin{array}{cr}
\hat{e}^c_k \\
\hat{e}^l_k 
\end{array}\right]
-q_v v_k^p + \vec{u}_k^\top R \vec{u}_k
\end{equation}
where $Q, q_v, R$ are the respective weights. We extend the formulation of \cite{liniger} to account for moving obstacles and lanes using the Potential Field method \cite{siebenrock}. 
\begin{equation}\label{eqn5_7}
    \begin{split}
    \centering
    J^p_{k} &= q_{ob} \sum_{i=1}^{O} \cdot \exp\left(-\left(\frac{\Delta x^o_k}{l^o}\right)^2 - \left(\frac{\Delta y^o_k}{w^o}\right)^2\right) \\
    & + q_{lm} \sum_{l=0}^{L} \exp\left(-\left(\frac{d_{lm}^l-\hat{e}^c_k\left(\theta_k\right)}{\sigma}\right)^2\right)
    \end{split} 
\end{equation}
where $\Delta x^i_k, \Delta y^i_k$ are the distances to the respective obstacle, $d_{lm}^l$ is the signed distance from the reference path to the respective $L$ lane marker, $\sigma$ a scaling factor and $l^i, w^i$ a conservative estimation of the length and width of the obstacle and $q_{ob}, q_{lm}$ are the respective weights. Additionally, for the hard constraints we employ ellipses to approximate the occupied area by the obstacles and utilize a union of three circles to approximate the ego's occupied space. With that, we approximate the Minkowsky sum as described in \cite{brito1}. The trajectories of the obstacles are provided by the prediction module which will be introduced in the next section.  
To ensure that the AV stays within the road boundaries, we impose linear constraints
\begin{equation}
    -d_{lb}\left(\theta_k\right) \leq \hat{e}^c_k\left(\theta_k\right) \leq d_{rb}\left(\theta_k\right).
\end{equation}

Box constraints are imposed on the control inputs $j_k \in [j_{\text{min}}, j_{\text{max}}]$ and $\dot{\delta}_k \in [\dot{\delta}_{\text{min}}, \dot{\delta}_{\text{max}}]$. Additionally, we limit $\delta$, $a$, and the lateral acceleration to ensure that the trajectories are feasible for the vehicle\cite{polack}.
This leaves us with the nonconvex optimization problem:
\begin{equation}
\label{eqn:J}
\begin{aligned}
 \underset{\vec{Z},\vec{U}}{\min} \ & \sum_{k=0}^{N-1} (J_k(\vec{z}_k, \vec{u}_k) + J^p_k(\vec{z}_k)) + J_N(\vec{z}_N) 
 \end{aligned}
\end{equation}

\begin{equation}
\begin{aligned}
 & s.t.  \   \vec{z}_{k+1} = f(\vec{z}_{k},\vec{u}_{k}^H) \\
 & \vec{z}_0 = \vec{z}(0) \\
 & \vec{z}_{k} \in  \mathcal{Z}, \ \vec{u}_{k} \in \mathcal{U} \nonumber
\end{aligned}
\end{equation}
where $\mathcal{Z}$ is set of state constraints imposed by road boundaries, obstacles, and lateral acceleration,  $\mathcal{U}$ is the set of box constraints on the control inputs and  $J_N(\vec{z}_N) $ is the terminal cost.
The trajectories planned by the MPC are denoted by $\tau = [\vec{Z},\vec{U}]^\top $ where $\vec{Z} = [\vec{z}_0, ..., \vec{z}_N]^\top$ and $\vec{U} = [\vec{u}_0, ..., \vec{u}_{N-1}]^\top$ where $N$ is the prediction steps. 


\section{Learning-aided Warmstart} \label{sec:warmstart}
\begin{figure*}[t]
\vspace{0.6em} 
    \centering
    \includegraphics[width=1\textwidth]{"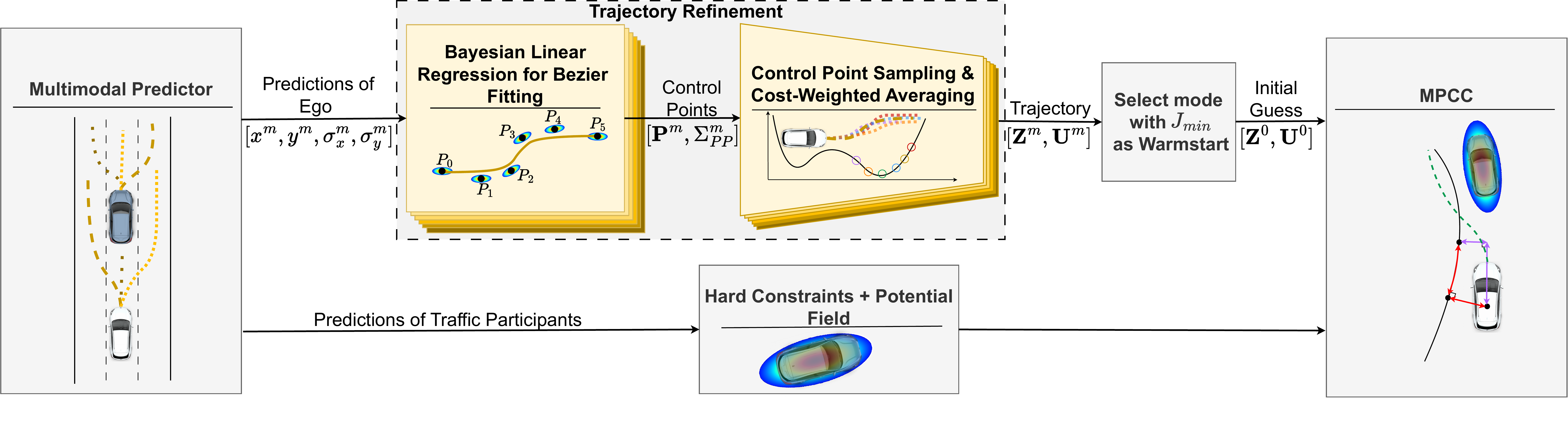"}
    \caption{The learning-aided warmstart framework  for motion planning and control using a multimodal predictor}
    \label{fig:method}
    \vspace{-1.0em} 
\end{figure*}
This section introduces our learning-aided approach, which we append as a warmstart method to the baseline MPCC introduced in sec. \ref{sec:MPCC} (s. Fig. \ref{fig:method}). The warmstart aims to provide an initial guess $\vec{\tau}^0 = [ \vec{Z}^0, \vec{U}^0 ]$ sufficiently close to a satisfactory local optimum of the current time $t_k$,  $||\vec{\tau}^0- \vec{\tau}^*_k||\leq \vec{\epsilon}$ such that the MPCC optimizer locally converges to this optimum. The conventional approach of warmstarting using the optimal trajectory of the last timestep $\vec{\tau}^*_{k-1} =[ \vec{Z}^*_{k-1}, \vec{U}^*_{k-1} ]$ is still employed in our framework by comparing it to the learning-aided output in terms of minimum cost at timestep $k$ i.e., considering the new information e.g., about obstacle motion. This allows for enhancing the convergence quality while still maintaining an upper bound for the cost provided by $\vec{\tau}^*_{k-1}$.
\subsection{Motion Predictor for Trajectory Proposals} \label{subsec:MTR}

Motion prediction models reason about the map, the historical trajectories of objects, and their interactions to forecast objects' future movement. However, determining the intentions of other traffic participants considering the various choices an agent can make (e.g. whether a car will overtake or follow a leading vehicle) is challenging. 
To address this challenge, many learning-based motion prediction models opt to provide \emph{multimodal} predictions. Motion Transformer (MTR) \cite{shi_motion_2022} and Wayformer \cite{nayakanti_wayformer_2022},  are examples of such multi-modal predictors trained on large-scale motion prediction datasets, such as Waymo Open Motion (WO) \cite{ettinger_waymo_2021}.  
In our method, we employ MTR which outputs a Gaussian Mixture Model (GMM) for the object's future position $\mathcal{N}(\vec{\mu}^m, \boldsymbol{\Sigma}^m_{in})$ at every timestep. Each component $m \in  \mathcal{M}$ of this mixture corresponds to one predicted mode.

We capitalize on the necessity of the predictor for obstacle prediction\footnote{In this study, we use the most probable obstacle prediction. Planning with multimodal obstacle predictions remains future work.} and reutilize it for predicting the ego trajectory. The aim of our approach is to leverage the multimodal output of the predictor to identify multiple local optima and select the best one. To elaborate on that, we introduce the concept of homotopy classes in the context of motion planning \cite{bhatta}.
\begin{definition}
 Two trajectories connecting the same start and end position belong to the same homotopy class if they can be continuously
deformed into each other without intersecting an obstacle. The set of all trajectories that are homotopic to each other is denoted as homotopy class.
\end{definition}
According to Definition 1, homotopic trajectories share the same start and end point. Due to the initial state constraint, the trajectories always share the same start by definition. However, we relax the end point requirement as suggested in \cite{bender}. 
With the assumption that obstacles are the cause for the existence of multiple minima, it follows that all trajectories $\vec{\tau}_i \in \mathcal{A}$ are homotopic, where $\mathcal{A} = \{\vec{\tau} \in \mathcal{X}| J(\vec{\tau}^*) \leq J(\vec{\tau}_i), \vec{\tau}_i = \vec{\tau}^* + \vec{\epsilon}, \vec{\epsilon} \geq 0   \}$ denotes the attractive vicinity of the local optimizer $\vec{\tau}^*$\cite{rosmann}.

Thus, the aim of the learning-aided warmstart can be further specified in providing an initial trajectory from the right homotopy class. One of the main causes of multimodality in motion prediction is the interaction with other traffic participants. Consequently, different modes often correspond to different homotopy classes. Therefore, we make the following assumptions:
\begin{assumption}
Several of the predicted modes do not share the same homotopy class and cover a subset of the existing homotopy classes $h \in \mathcal{H}$,  i.e. $|\{ [m]| m \in \mathcal{M} \} \cap \mathcal{H}| \geq 2 $.  
\end{assumption}

\begin{assumption}
The covariance of the components of the GMMs i.e. for the respective mode is small enough such that trajectories drawn from the same components correspond to the same homotopy class (s. Fig. \ref{fig:P3}). 
\end{assumption}

Subsequently, we introduce how to utilize and further refine these provided modes to be able to select the best one (in terms of cost) as a warmstart.

\subsection{Bezier Curve Fitting} \label{subsec:Bayes}
Typical trajectory predictors such as MTR predict only the object position distributions 
at every prediction timestamp. However, we require the complete state and control input trajectories for our warmstart. Furthermore, our method is required to sample realistic trajectories from the given distribution to further refine the predictor trajectory.

We select 5th-degree Bezier Curves to fit the predicted positions. They represent the optimal solutions in terms of travel time, control effort, and jerk \cite{werling} and thus are close to the optimal vehicle trajectories outputted by the MPC. This smooths the often jerky predictions and allows us to calculate derivatives analytically. Further, we can perform this fit in such a way as to match current kinematic state. This continuity constraint is not directly enforced by MTR. 
We perform the fitting using Bayesian Linear Regression (BLR) to output a distribution over the curve parameters from which we can sample in the next step.
 The 5th-degree Bezier Curve $\vec{c}(t)$ can be expressed as a linear combination of 6  control points $\vec{P}_j \in \mathbb{R}^2 $ and the Bernstein polynomials  $\phi_j(t) : 	\mathbb{R} \to 	\mathbb{R}$:
\begin{equation}\label{eq:c}
    \vec{c}(t) = \sum^5_{j=0} \phi_j(t) \vec{P}_j
\end{equation}
From the temporal derivatives of the Bezier curve, we can then calculate the state and control input trajectories.

Hence, we first need to estimate the control points $\vec{P}^m_j$ from the output of the predictor for each mode. The initial guess should ideally satisfy the continuity constraint  $\vec{z}_0 = \vec{z}(0)$. For this, we exploit the property of the Bezier curve that the initial conditions can be determined from the control points. The initial condition for the  $(d^{\textrm{th}})$ derivative can be calculated as:

\begin{equation}
    \vec{c}^{(d)}(0) = \frac{6!}{(6-d)!}\Delta^d \vec{P}_0
\end{equation}
where $\Delta^k$ is the forward difference operator recursively defined by
$
\Delta^k \vec{P}_i=\Delta^{k-1} \vec{P}_{i+1}-\Delta^{k-1} \vec{P}_i \quad \text { where } \quad \Delta^0 \vec{P}_i=\vec{P}_i.
$
From this, we determine the first three control points:

\begin{align}\label{eq:c0}
 \vec{c}(0) = & [x_0, y_0]^\top, \ \ \vec{\dot{c}}(0) = [v_0 \cos(\psi_0), v_0 \sin(\psi_0)]^\top,  \nonumber \\
 \vec{\ddot{c}}(0) = & [a_0 \cos(\psi_0) - \frac{v_0^2}{l} \tan(\delta_0) \sin(\psi_0), \nonumber\\   &a_0 \sin(\psi_0) - \frac{v_0^2}{l} \tan(\delta_0) \cos(\psi_0)]^\top    
\end{align}
We utilize these relationships for the control points as a strong Gaussian prior $\mathcal{N}(\vec{P}^{m,0}, \vec{\Sigma}^{m,0})$ for the BLR. In other words, the first three elements of $\vec{P}^{m,0}$ correspond to eq. \ref{eq:c0}, and $\vec{\Sigma}^{m,0}$ are derived from the tracked uncertainty of the states from the on-board sensors. As for the remaining three elements in $\vec{P}^{m,0}$, we employ an uninformed prior.

To formulate the BLR-problems we form $\mathbf{C}^m=\boldsymbol{\Phi}^{\mathrm{T}} \vec{P}^m$ with the vector of the  control points $\vec{P}^m \in \mathbb{R}^{6 \cdot 2}$, the vector of Bezier curve points $\vec{C}^m \in \mathbb{R}^{2N}$ and the new basis function $\boldsymbol{\Phi}\in \mathbb{R}^{6 \cdot 2 \times 2N}$ as done in  \cite{yao, reichardt}.
The uncertainties of the predictions $[\vec{X}_m, \vec{Y}_m]^\top$, i.e. the covariance $\vec{\Sigma}^m_{in}$ outputted by the GMM enter as Gaussian observation noise  into the regression.
\begin{equation}
 [\vec{X}_m, \vec{Y}_m]^\top  =  \vec{\Phi}^\top \vec{P}^m + \vec{e}, \ \vec{e} \sim N(0, \vec{\Sigma}^m_{in} )
\end{equation}
Consequently, the posterior and the covariance for the control points are given:
\begin{align}\label{eq:gm}
 \vec{\Sigma}^m_{PP} = & \left(\vec{\Phi} \vec{\Sigma}^m_{in} \vec{\Phi}^\top +
 \left(\vec{\Sigma}^{m,0}\right)^{-1}  \right)^{-1}   \\
 \vec{P^m} = & \vec{\Sigma}^m_{PP} \vec{\Phi}^\top \left(\vec{\Sigma}^m_{in}\right)^{-1} 
 \left[\begin{array}{cr}
\vec{X}_m \\
\vec{Y}_m 
\end{array}\right]
 +  \vec{\Sigma}^m_{PP} \left(\vec{\Sigma}^{m,0}\right)^{-1} \vec{P}^{m,0}  \nonumber
\end{align}

\subsection{Control Point Sampling and Cost-Weighted Averaging } 
\label{subsec:Sampling}
Simply calculating the states and control inputs from each outputted modes of the predictor leads often to an ineffective warmstart. Even if the best homotopy class is chosen from these modes, it can still lead to a solution far from the optimum, resulting in a prolonged convergence time. Additionally, comparing modes to select the best homotopy class based on the cost function is inaccurate, as for two trajectories in two different homotopy classes $\vec{\tau}_{h,1}, \vec{\tau}_{h,2}$  $J(\vec{\tau}_{h,1}) > J(\vec{\tau}_{h,2})$ does not necessarily imply $J(\vec{\tau}^*_{h,1}) > J(\vec{\tau}^*_{h,2})$ since $\vec{\tau}_{h,1}$ and $\vec{\tau}_{h,2}$ can have different distances from their local optimum $\vec{\tau}^*_{h,1}$ and $\vec{\tau}^*_{h,2}$.
\begin{figure}[h]
\centering
\vspace{0.3em}
    \includegraphics[width=0.47\textwidth]{"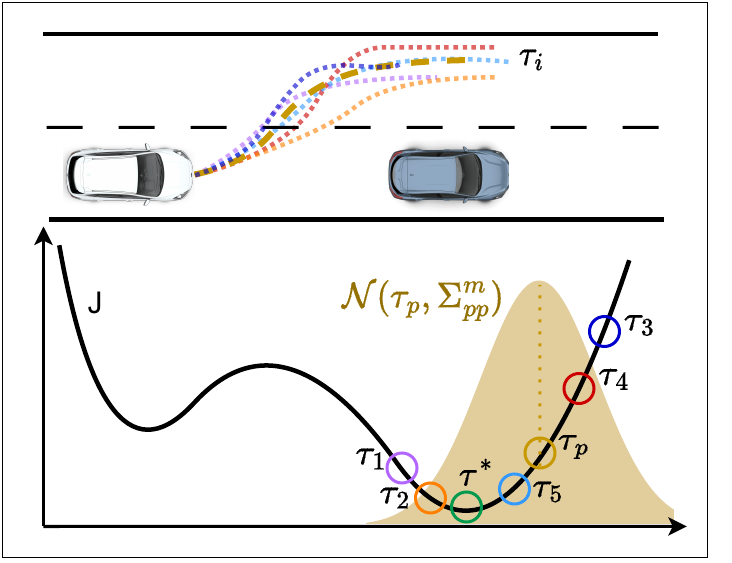"}
    \caption{Intuition of the trajectory sampling around the fitted trajectory $\vec{\tau}_{p}$ of the predictor which is assumed to fall in the attractive vicinity of local minimum}
    \label{fig:P3}
    \vspace{-1.3em} 
\end{figure}
Our solution to refine the trajectories is to utilize the distributions $\mathcal{N}(\vec{P}^m, \vec{\Sigma}_{pp}^m)$ from eqn.  \ref{eq:gm}, in order to sample $S$ Bezier curves (s. Fig. \ref{fig:P3} ). For each sample $\vec{P}_s^m$ and the mean $\vec{P}^m$, we compute the cost function and obtain our output control points through a cost-weighted average.

\begin{equation}\label{eq:wa}
\Bar{\vec{P}}^m = \sum_{s=0}^S w_s \Tilde{\vec{P}}_s^m \ \textrm{with} \ w_s = \frac{\textrm{e}^{-\lambda J(\Tilde{\vec{P}}_s^m)}}{\sum_{i=0}^S \textrm{e}^{-\lambda J(\Tilde{\vec{P}}_i^m)}}
\end{equation}
where $\lambda$ is a tunable parameter. The chosen weighting factor $w_n$ has the advantageous property that assigns significantly less influence to samples with markedly higher cost function compared to the sample with the minimum cost, i.e., $J(\Tilde{\vec{P}}_i^m) >> J(\Tilde{\vec{P}}_j^m), w_i \rightarrow 0$ (softmin normalization \cite{houghton}). This implies that, in essence, we are only giving substantial consideration to a limited subset of samples within a similar cost range.

Furthermore, recall Assumption 2, i.e.; consequently, we assume the samples are in the attractive vicinity of a local minimum. Provided the samples are well distributed around the region of convexity of this local optimum, taking the weighted average of the trajectories gives us a value inside the area spanned by the sample points. Hence, the output results generally in a trajectory closer to the minimum. 

We execute this process for each mode. Subsequently, the costs for each mode are compared, and the best one is employed as the warmstart. While this still does not guarantee the selection of the homotopy class of the global optimum, it allows us to choose a satisfactory local minimum at least. In autonomous driving, various maneuvers are often similarly satisfactory, and only undesired local minima must be prevented.

The detailed steps of the trajectory refinement are outlined in Algorithm 1.
It is important to note that this approach supports parallel computation due to the parallel nature of sampling and the independence of each trajectory from one another, i.e., it can take advantage of the parallel processing capabilities of modern GPUs, making it highly efficient.

\begin{algorithm}[ht]
\caption{Learning-aided Warmstart}
\SetAlgoLined
\SetKwInOut{Input}{Input}
\SetKwInOut{Output}{Output}
\Input{Measured state values $ \vec{z}_k = [x_k, y_k, \psi_k, v_k, a_{k}, \delta_k, \theta_k]^\top$, \\
optimal traj.last timestep $\vec{Z}^*_{k-1}$, $\vec{U}^*_{k-1}$,\\
map information $\vec{M_{r}}$,
reference path $\mathcal{P}_{ref}$, \\
pose history $ \vec{\eta}^k_{o}$  $\forall$ Agents $o$ with $o=0$ \\ denoting the ego vehicle
}
\Output{Initial Guess for MPCC $\vec{Z}^0$, $\vec{U}^0$}
    \textbf{Initialize} \#  refinement samples $S$, \# used modes $M$\\ 
\textit{//} Predict trajectories $\vec{\xi}^m_o = [\vec{x}, \vec{y}, \vec{\sigma}_x, \vec{\sigma}_y $]  \\
$\vec{\xi} \gets$  $\mathcal{MTR}(\vec{\eta}^k,\vec{M_{r}})$ \\
\ForEach{$m = \{1, ..., M\}$}{
    \textit{//} Fit predicted ego trajs. into Bezier Curve\\
    $ \vec{P}^m, \vec{\Sigma}_{pp}^m \gets$ BayesReg$(\vec{\xi}^m_0, \vec{z}_k)$  from \eqref{eq:gm} \\
    \textit{//} Sample from $\mathcal{N}(\vec{P}^m, \vec{\Sigma}_{pp}^m)$\\
    $\Tilde{\vec{P}}_1^m,\ldots,\Tilde{\vec{P}}_S^m \sim \mathcal{N}(\vec{P}^m, \vec{\Sigma}_{pp}^m)$\\
    \ForEach{$s = \{1, ..., S\}$}{
    \textit{//}  Calculate States, Control Inputs and Cost\\
    $\vec{Z}^{m}_s,\vec{U}^{m}_s, J^{m}_s  \gets  \mathcal{J}(\Tilde{\vec{P}}_s^m, \mathcal{P}_{ref}, \vec{\xi}^0_{1:O})$   \\
    }
    //  Calculate Cost-weighted Average of samples\\ 
    $\Bar{\vec{P}}^m \gets $ from \eqref{eq:wa}\\
    \textit{//}   Calculate States, Control Inputs and Cost \\
    $\vec{Z}^{m},\vec{U}^{m}, J^{m}  \gets  \mathcal{J}(\vec{\Bar{\vec{P}}^m}, \mathcal{P}_{ref}, \vec{\xi}^0_{1:O})$   \\
    } 
\textit{//} Select mode with minimal cost\\
$m^* \gets \argmin_{m} J^{m}$\\
\If{ $J^{m^*} \leq  \textrm{Cost of } \vec{Z}^*_{k-1}, \vec{U}^*_{k-1} \textrm{ at timestep } k$ }{
    \Return{($\vec{Z}^0 \gets \vec{Z}^{m^*}, \vec{U}^0 \gets \vec{U}^{m^*}$)}
}
\Else{
    \Return{($\vec{Z}^0 \gets \vec{Z}^*_{k-1}, \vec{U}^0 \gets \vec{U}^*_{k-1}$)}
} 
\end{algorithm}

\begin{table*}[]
    \vspace{0.5em} 
    \caption{Results of experiment III. Comparison of Baseline and the learning-aided Framework using Monte Carlo analysis}
    \vspace{-0.6em} 
    \centering
    \begin{tabular}{ |p{2.1cm}||c|c|c|c|c|c|c|c|  }
 \hline
  & \multicolumn{3}{c|}{\textbf{Merging Execution}} & \multicolumn{4}{c|}{\textbf{Convergence Quality}} & \multicolumn{1}{c|}{\textbf{}} \\ 
 \cline{2-9}
  & Success & Aborted & Collision & Success & \parbox{1.6cm}{Max. Time\\ exceeded} & \parbox{1.6cm}{Converge to\\ Infeasibility} & \parbox{1.6cm}{Average Cost} &  Average Solving time (std) \\ 
 \hline
 \parbox{2.1cm}{\textbf{Baseline MPCC}} &73 \% & 11\% & 16\% &  69.3\% & 13.6\% & 17.1\% & 4995 &   106 ms (40 ms)\\ 
  \hline
 \parbox{2.1cm}{\textbf{Our Framework}} & \textbf{88 \%} & \textbf{4\%}  & \textbf{8\%} & \textbf{82.7\%} & \textbf{7.0\%} & \textbf{10.3\%}  & \textbf{3737} & \textbf{94 ms (33 ms)}\\ 
 \hline
\end{tabular}
\label{tab:mca}
\vspace{-1.0em} 
\end{table*}


\section{Performance Evaluation} \label{sec:Results}
\begin{figure}[h]
\centering
    \includegraphics[trim = 26mm 0mm 5mm 10.5mm, clip, width=0.48\textwidth]{"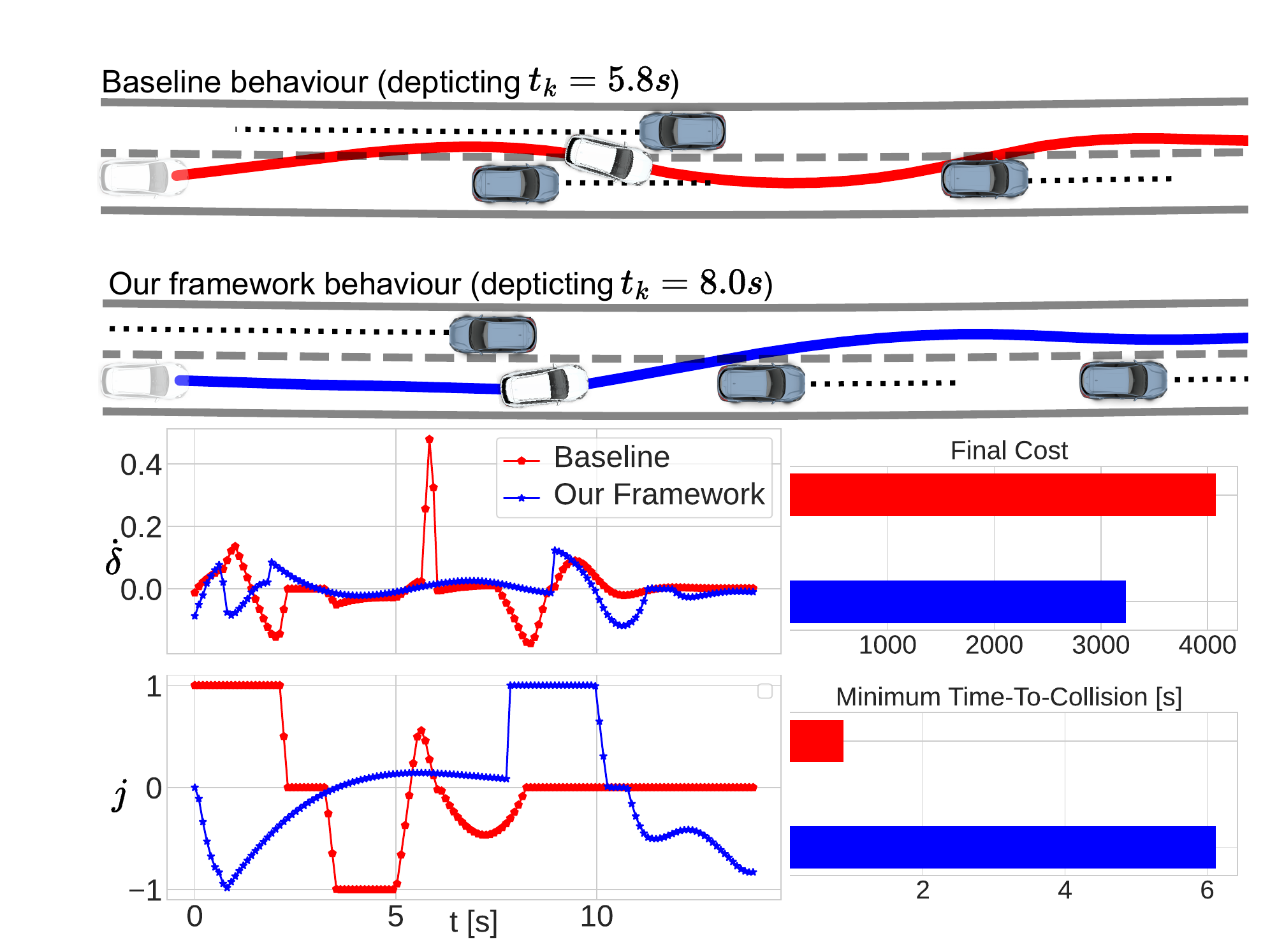"}
    \caption{Illustration of experiment I. Comparison of the local minimum that the baseline converged to with  our framework}
    \label{fig:P5}
    \vspace{-1.0em} 
\end{figure}
\begin{figure}[h]
\centering
    \includegraphics[trim = 10mm 0mm 0mm 0mm, clip, width=0.5\textwidth]{"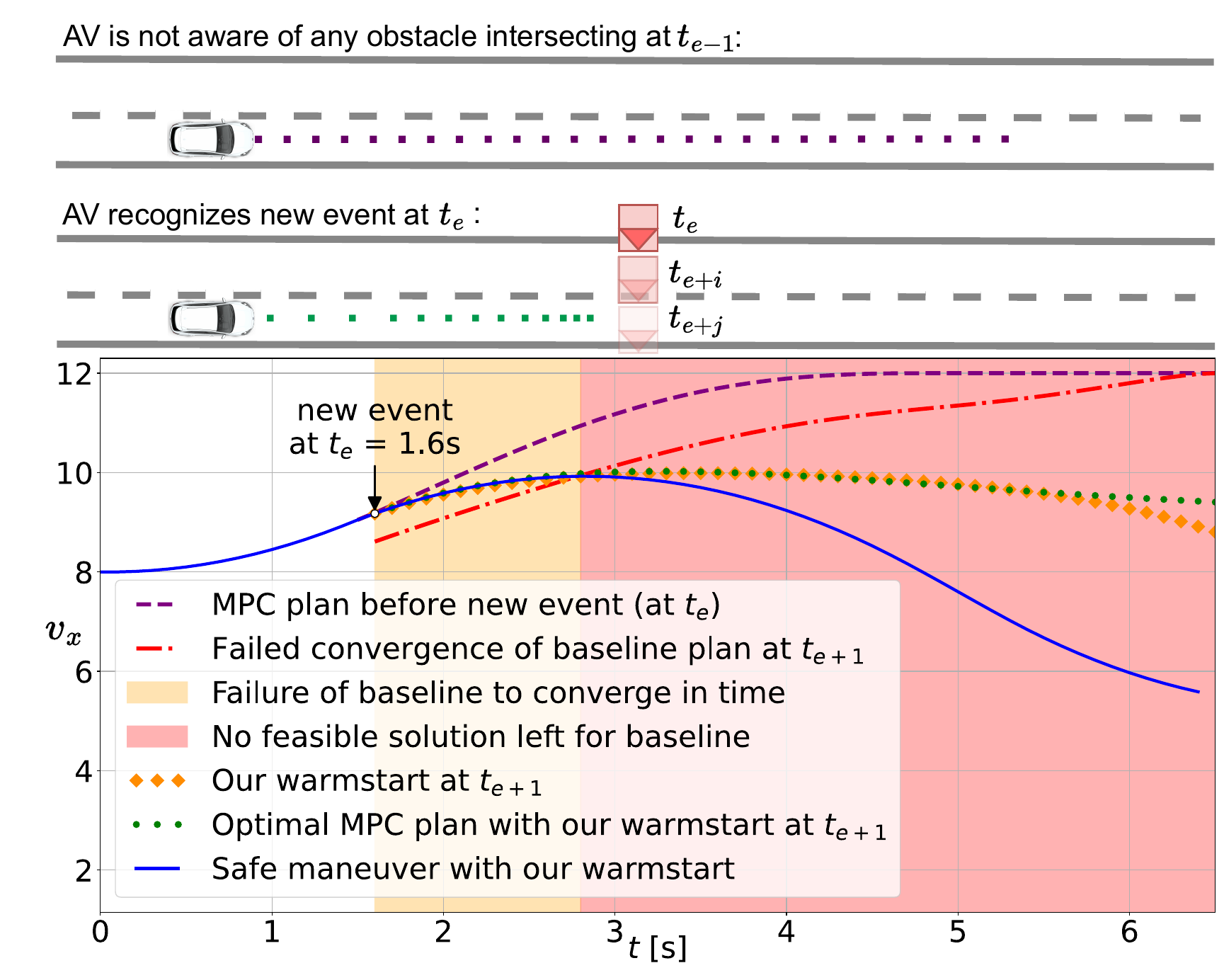"}
    \caption{Illustration of experiment II. Comparison of the baseline to our framework when an event occurs that changes the optimization problem between two timesteps}
    \label{fig:P4} 
    \vspace{-1.3em} 
\end{figure}
We compare our MPCC with learning-aided warmstart to the baseline MPCC with conventional warmstart in three experiments. The first two experiments serve as illustrative examples to highlight the two strengths of our approach. The last experiment entails a Monte Carlo simulation of random highway merging scenarios to provide statistical results (s. Tab. \ref{tab:mca}).

Experiment $I$ involves a scenario with two lanes, where the left lane  accommodates oncoming traffic but allows for overtaking. This scenario is well-suited to showcase the capability of our framework in escaping undesired local minima, as the presence of other traffic participants introduces non-convexity to the optimization problem. In \cite{bender, yi}, it is demonstrated that a planner in this scenario may converge towards several distinct local minima/homotopy classes. In our case, our learning-aided warmstart leads to a different behavior compared to the MPCC without warmstart (s. Fig. \ref{fig:P5}); i.e., the two planners converge towards two distinct local optima. The costs for our planner are significantly lower than those for the baseline planner (s. Fig. \ref{fig:P5}). This figure also provides a comparison of the control input trajectories and the minimum time-to-collision (TTC) for both planners in this scene, displaying the shortcomings of the local minima the baseline converged to.

Experiment II involves a scenario where an obstacle crosses the path of the ego vehicle. Initially, the ego vehicle is unaware of this occurrence for the first few moments, causing a sudden shift in the optimization problem for the planner. Such a situation can arise in various scenarios, for instance, when the obstacle is initially occluded or when predictions change (due to unknown intentions of traffic participants or a new decision of an object). In this case, the planner must be capable of finding a solution for the new optimization problem in real-time, even though it differs distinctly from the last timestep. Hence, we impose a maximum solving time constraint. However, we set this limit relatively high with \(t_{\text{max}} = 0.5s\) since there is potential to accelerate the MPC runtime through alternative implementations and hardware enhancements, etc. Despite this high maximum solving time, the baseline planner is unable to converge in time when the change occurs (from the new event at \(t_e = 1.6\) in Fig. \ref{fig:P4}). The red curve depicts the velocity trajectory outputted by the solver. This trajectory fails to satisfy both collision constraints and the initial condition. In such cases, it is customary to utilize the solution from the last time step, which, in this example, leads to further acceleration of the ego. This behavior ultimately results in an unavoidable collision. In contrast, Fig. \ref{fig:P4} depicts that our approach can directly provide an appropriate warmstart after the event.

For experiment III, we consider a highway merging scenario (s. Fig.~\ref{fig:P1}) and utilize the Intelligent Driver Model \cite{treiber} to simulate the behavior of the traffic participants. We generate 100 test runs by randomly sampling the parameters of the IDM model (such as desired velocity, minimum headway, etc.), as well as the initial positions and velocities for the ego vehicle and the other vehicles. As a result, we compare the rate of successful mergings, the percentage of the ego getting stuck in the entrance lane, and collisions. Additionally, we assess the convergence quality in terms of the percentage of successful convergence, failed convergence due to reaching the time limit, and failed convergence due to converging to a point of infeasibility. Further benchmarking parameters are the average cost and solving time\footnote{Solving time data is for comparative purposes only and should not be taken as absolute}. The significant performance improvement to the baseline becomes evident when considering highway merging. Firstly, each gap between traffic participants potentially corresponds to a local minimum where one can clearly be better than the other e.g., due to gap size. Secondly, shifts in motion predictions of the traffic participants during merging often substantially impact the ego vehicle's plan e.g., if a prediction changes the acceleration slightly, the optimal plan for the ego may shift from merging in front to merging behind.


\section{CONCLUSIONS}\label{sec:conclusion}

A Learning-aided Warmstart Framework is proposed to address the problem of Model Predictive Control with local minima and convergence issues if using the conventional warmstart strategy in fast-changing, uncertain environments. This framework leverages a multimodal predictor that predicts trajectories for traffic participants and the ego vehicle, respectively. The different ego trajectory modes are used to identify multiple homotopy classes, each associated with an attractive vicinity of a different local optimum. To achieve this, we introduced a novel sampling-based trajectory refinement approach using Bayesian Linear Regression for Bezier Curve Fitting to efficiently optimize the trajectories before selecting the best one as an initial guess. Our Monte Carlo analysis demonstrates that our framework significantly improves the convergence quality in highway merging scenarios.

\bibliographystyle{IEEEtran}
\bibliography{Refs}

\begin{thebibliography}{10}
\providecommand{\url}[1]{#1}
\csname url@samestyle\endcsname
\providecommand{\newblock}{\relax}
\providecommand{\bibinfo}[2]{#2}
\providecommand{\BIBentrySTDinterwordspacing}{\spaceskip=0pt\relax}
\providecommand{\BIBentryALTinterwordstretchfactor}{4}
\providecommand{\BIBentryALTinterwordspacing}{\spaceskip=\fontdimen2\font plus
\BIBentryALTinterwordstretchfactor\fontdimen3\font minus \fontdimen4\font\relax}
\providecommand{\BIBforeignlanguage}[2]{{%
\expandafter\ifx\csname l@#1\endcsname\relax
\typeout{** WARNING: IEEEtran.bst: No hyphenation pattern has been}%
\typeout{** loaded for the language `#1'. Using the pattern for}%
\typeout{** the default language instead.}%
\else
\language=\csname l@#1\endcsname
\fi
#2}}
\providecommand{\BIBdecl}{\relax}
\BIBdecl

\bibitem{micheli}
F.~Micheli, M.~Bersani, S.~Arrigoni, F.~Braghin, and F.~Cheli, ``{NMPC} trajectory planner for urban autonomous driving,'' \emph{Vehicle system dynamics}, vol.~61, no.~5, pp. 1387--1409, 2023.

\bibitem{siebenrock}
F.~Siebenrock, M.~G{\"u}nther, and S.~Hohmann, ``{LTV}-{MPC} {Based} {Trajectory} {Planning} {Considering} {Uncertain} {Object} {Prediction} {Through} {Adaptive} {Potential} {Fields},'' in \emph{2020 IEEE Conference on Control Technology and Applications (CCTA)}.\hskip 1em plus 0.5em minus 0.4em\relax IEEE, 2020, pp. 666--672.

\bibitem{shi}
Q.~Shi, J.~Zhao, A.~El~Kamel, and I.~Lopez-Juarez, ``{MPC} {Based} {Vehicular} {Trajectory} {Planning} in {Structured} {Environment},'' \emph{IEEE Access}, vol.~9, pp. 21\,998--22\,013, 2021.

\bibitem{liniger}
A.~Liniger, A.~Domahidi, and M.~Morari, ``Optimization-{Based} {Autonomous} {Racing} of 1:43 {Scale} {RC} {Cars},'' \emph{Optimal Control Applications and Methods}, vol.~36, no.~5, pp. 628--647, 2015.

\bibitem{brito1}
B.~Brito, B.~Floor, L.~Ferranti, and J.~Alonso-Mora, ``Model {Predictive} {Contouring} {Control} for {Collision} {Avoidance} in {Unstructured} {Dynamic} {Environments},'' \emph{IEEE Robotics and Automation Letters}, vol.~4, no.~4, pp. 4459--4466, 2019.

\bibitem{lyons}
L.~Lyons and L.~Ferranti, ``Curvature-{Aware} {Model} {Predictive} {Contouring} {Control},'' in \emph{2023 IEEE International Conference on Robotics and Automation (ICRA)}, 2023, pp. 3204--3210.

\bibitem{aradi}
S.~Aradi, ``Survey of deep reinforcement learning for motion planning of autonomous vehicles,'' \emph{IEEE Transactions on Intelligent Transportation Systems}, vol.~23, no.~2, pp. 740--759, 2022.

\bibitem{kendall}
A.~Kendall, J.~Hawke, D.~Janz, P.~Mazur, D.~Reda, J.-M. Allen, V.-D. Lam, A.~Bewley, and A.~Shah, ``Learning to drive in a day,'' 2018.

\bibitem{gros}
S.~Gros and M.~Zanon, ``Data-driven {Economic} {NMPC} using {Reinforcement} {Learning},'' \emph{IEEE Transactions on Automatic Control}, vol.~65, no.~2, pp. 636--648, 2019.

\bibitem{zarrouki}
B.~Zarrouki, V.~Klös, N.~Heppner, S.~Schwan, R.~Ritschel, and R.~Voßwinkel, ``Weights-varying {MPC} for {Autonomous} {Vehicle} {Guidance}: a {Deep} {Reinforcement} {Learning} {Approach},'' in \emph{2021 {European} {Control} {Conference} ({ECC})}, Jun. 2021, pp. 119--125.

\bibitem{brito2}
B.~Brito, M.~Everett, J.~P. How, and J.~Alonso-Mora, ``Where to go {Next}: {Learning} a {Subgoal} {Recommendation} {Policy} for {Navigation} in {Dynamic} {Environments},'' \emph{IEEE Robotics and Automation Letters}, vol.~6, no.~3, pp. 4616--4623, 2021.

\bibitem{tamar}
A.~Tamar, G.~Thomas, T.~Zhang, S.~Levine, and P.~Abbeel, ``Learning from the {Hindsight} {Plan} -- {Episodic} {MPC} {Improvement},'' in \emph{2017 IEEE International Conference on Robotics and Automation (ICRA)}.\hskip 1em plus 0.5em minus 0.4em\relax IEEE, 2017, pp. 336--343.

\bibitem{Koller}
T.~Koller, F.~Berkenkamp, M.~Turchetta, J.~B{\"o}decker, and A.~Krause, ``Learning-based model predictive control for safe reinforcement learning,'' in \emph{Robust autonomy: tools for safety in real-world uncertain environments}, 2019.

\bibitem{Berberich}
J.~Berberich, J.~K{\"o}hler, M.~A. M{\"u}ller, and F.~Allg{\"o}wer, ``Data-{Driven} {Model} {Predictive} {Control} {With} {Stability} and {Robustness} {Guarantees},'' \emph{IEEE Transactions on Automatic Control}, vol.~66, no.~4, pp. 1702--1717, 2020.

\bibitem{Hewing}
L.~Hewing, K.~P. Wabersich, M.~Menner, and M.~N. Zeilinger, ``Learning-{Based} {Model} {Predictive} {Control}: {Toward} {Safe} {Learning} in {Control},'' \emph{Annual Review of Control, Robotics, and Autonomous Systems}, vol.~3, pp. 269--296, 2020.

\bibitem{brito3}
B.~Brito, A.~Agarwal, and J.~Alonso-Mora, ``Learning {Interaction}-aware {Guidance} {Policies} for {Motion} {Planning} in {Dense} {Traffic} {Scenarios},'' \emph{arXiv preprint arXiv:2107.04538}, 2021.

\bibitem{song}
Y.~Song and D.~Scaramuzza, ``Policy {Search} for {Model} {Predictive} {Control} with {Application} to {Agile} {Drone} {Flight},'' \emph{IEEE Transactions on Robotics}, vol.~38, no.~4, pp. 2114--2130, 2022.

\bibitem{Wabersich}
K.~P. Wabersich and M.~N. Zeilinger, ``A predictive safety filter for learning-based control of constrained nonlinear dynamical systems,'' \emph{Automatica}, vol. 129, p. 109597, 2021.

\bibitem{Tearle}
B.~Tearle, K.~P. Wabersich, A.~Carron, and M.~N. Zeilinger, ``A predictive safety filter for learning-based racing control,'' \emph{IEEE Robotics and Automation Letters}, vol.~6, no.~4, pp. 7635--7642, 2021.

\bibitem{mansard}
N.~Mansard, A.~DelPrete, M.~Geisert, S.~Tonneau, and O.~Stasse, ``Using a {Memory} of {Motion} to {Efficiently} {Warm}-{Start} a {Nonlinear} {Predictive} {Controller},'' in \emph{2018 IEEE International Conference on Robotics and Automation (ICRA)}.\hskip 1em plus 0.5em minus 0.4em\relax IEEE, 2018, pp. 2986--2993.

\bibitem{kabutan}
T.~Barbi{\'e}, R.~Kabutan, R.~Tanaka, and T.~Nishida, ``Gaussian mixture spline trajectory: learning from a dataset, generating trajectories without one,'' \emph{Advanced Robotics}, vol.~32, no.~10, pp. 547--558, 2018.

\bibitem{natarajan}
S.~Natarajan, ``Learning initial trajectory using sequence-to-sequence approach to warm start an optimization-based motion planner,'' in \emph{2021 IEEE/RSJ International Conference on Intelligent Robots and Systems (IROS)}.\hskip 1em plus 0.5em minus 0.4em\relax IEEE, 2021, pp. 9430--9436.

\bibitem{lembono}
T.~S. Lembono, A.~Paolillo, E.~Pignat, and S.~Calinon, ``Memory of {Motion} for {Warm}-starting {Trajectory} {Optimization},'' \emph{IEEE Robotics and Automation Letters}, vol.~5, no.~2, pp. 2594--2601, 2020.

\bibitem{svenstrup}
M.~Svenstrup, T.~Bak, and H.~J. Andersen, ``Trajectory planning for robots in dynamic human environments,'' in \emph{2010 IEEE/RSJ International Conference on Intelligent Robots and Systems}.\hskip 1em plus 0.5em minus 0.4em\relax IEEE, 2010, pp. 4293--4298.

\bibitem{li}
J.~Li, M.~Ran, H.~Wang, and L.~Xie, ``{MPC}-based {Unified} {Trajectory} {Planning} and {Tracking} {Control} {Approach} for {Automated} {Guided} {Vehicles},'' in \emph{2019 IEEE 15th International Conference on Control and Automation (ICCA)}.\hskip 1em plus 0.5em minus 0.4em\relax IEEE, 2019, pp. 374--380.

\bibitem{degroot}
O.~de~Groot, L.~Ferranti, D.~Gavrila, and J.~Alonso-Mora, ``Globally {Guided} {Trajectory} {Planning} in {Dynamic} {Environments},'' in \emph{2023 IEEE International Conference on Robotics and Automation (ICRA)}.\hskip 1em plus 0.5em minus 0.4em\relax IEEE, 2023, pp. 10\,118--10\,124.

\bibitem{rosmann}
C.~R{\"o}smann, F.~Hoffmann, and T.~Bertram, ``Planning of multiple robot trajectories in distinctive topologies,'' in \emph{2015 European Conference on Mobile Robots (ECMR)}.\hskip 1em plus 0.5em minus 0.4em\relax IEEE, 2015, pp. 1--6.

\bibitem{yi}
B.~Yi, P.~Bender, F.~Bonarens, and C.~Stiller, ``Model {Predictive} {Trajectory} {Planning} for {Automated} {Driving},'' \emph{IEEE Transactions on Intelligent Vehicles}, vol.~4, no.~1, pp. 24--38, 2018.

\bibitem{polack}
P.~Polack, F.~Altch{\'e}, B.~d'Andr{\'e}a Novel, and A.~de~La~Fortelle, ``Guaranteeing {Consistency} in a {Motion} {Planning} and {Control} {Architecture} {Using} a {Kinematic} {Bicycle} {Model},'' in \emph{2018 Annual American Control Conference (ACC)}.\hskip 1em plus 0.5em minus 0.4em\relax IEEE, 2018, pp. 3981--3987.

\bibitem{shi_motion_2022}
S.~Shi, L.~Jiang, D.~Dai, and B.~Schiele, ``Motion transformer with global intention localization and local movement refinement,'' \emph{Advances in Neural Information Processing Systems}, vol.~35, pp. 6531--6543, 2022.

\bibitem{nayakanti_wayformer_2022}
N.~Nayakanti, R.~Al-Rfou, A.~Zhou, K.~Goel, K.~S. Refaat, and B.~Sapp, ``{Wayformer}: {Motion} {Forecasting} via {Simple} \& {Efficient} {Attention} {Networks},'' \emph{arXiv preprint arXiv:2207.05844}, 2022.

\bibitem{ettinger_waymo_2021}
S.~Ettinger, S.~Cheng, B.~Caine, C.~Liu, H.~Zhao, S.~Pradhan, Y.~Chai, B.~Sapp, C.~R. Qi, Y.~Zhou \emph{et~al.}, ``Large scale interactive motion forecasting for autonomous driving: The waymo open motion dataset,'' in \emph{Proceedings of the IEEE/CVF International Conference on Computer Vision}, 2021, pp. 9710--9719.

\bibitem{bhatta}
S.~Bhattacharya, \emph{Topological and geometric techniques in graph search-based robot planning}.\hskip 1em plus 0.5em minus 0.4em\relax University of Pennsylvania, 2012.

\bibitem{bender}
P.~Bender, {\"O}.~{\c{S}}. Ta{\c{s}}, J.~Ziegler, and C.~Stiller, ``The combinatorial aspect of motion planning: Maneuver variants in structured environments,'' in \emph{2015 IEEE Intelligent Vehicles Symposium (IV)}.\hskip 1em plus 0.5em minus 0.4em\relax IEEE, 2015, pp. 1386--1392.

\bibitem{werling}
M.~Werling, J.~Ziegler, S.~Kammel, and S.~Thrun, ``Optimal trajectory generation for dynamic street scenarios in a frenét frame,'' in \emph{2010 IEEE International Conference on Robotics and Automation}, 2010, pp. 987--993.

\bibitem{yao}
Y.~Yao, D.~Goehring, and J.~Reichardt, ``An empirical bayes analysis of vehicle trajectory models,'' \emph{arXiv preprint arXiv:2211.01696}, 2022.

\bibitem{reichardt}
J.~Reichardt, ``Trajectories as markov-states for long term traffic scene prediction,'' in \emph{14-th UniDAS FAS-Workshop}, Berkheim, Germany, 2022, p.~14.

\bibitem{houghton}
M.~D. Houghton, A.~B. Oshin, M.~J. Acheson, E.~A. Theodorou, and I.~M. Gregory, ``Path planning: Differential dynamic programming and model predictive path integral control on vtol aircraft,'' in \emph{AIAA SCITECH 2022 Forum}, 2022, p. 0624.

\bibitem{treiber}
M.~Treiber, A.~Hennecke, and D.~Helbing, ``Congested traffic states in empirical observations and microscopic simulations,'' \emph{Physical review E}, vol.~62, no.~2, p. 1805, 2000.

\end{thebibliography}
 
\end{document}